\documentclass[12pt]{article}
\usepackage[pctex32]{graphicx}
 
\begin {document}

\title{Status of e-Print Servers}
\author{Mauricio Ayala-S\'anchez\\
Departamento de Matem\'aticas\\
Universidad de los Andes, Bogot\'a, Colombia\\
E-mail: mau-ayal@uniandes.edu.co
}
\date{}
\maketitle

\abstract{We make a short study of the history and evolution of scientific publications, in order to explain the format for near-term e-Prints servers, proposing a new scientific publication scheme via digital network, and exploring the new dynamics of publication.}

\section {Introducci\'on}
Today's communication digital network (email, e-Print server, web sites, FTP, chat, etc.) has become a vital tool for the development of scientific research. Now it's possible to communicate with the scientific community in less than 24 hours any development or investigation that is considered of importance, in the same way we can access investigations and works published in the day. Additionally there are no restrictions to the size of a document, so it's possible to show a work with greater detail, even in an interactive way, at very low costs. Thus, the scientific communications have evolved from print. So, on one side it's not necessary to wait for months to read an article which is to be published in a journal, on the other hand it's not always possible to consult every journal printed anywhere of the world. 

\section {History and Evolution of Scientific Publications}
In the XVIIth century, the intellectual interchange occurred initially through personal letters, manuscripts or messengers. Science was being born and the scientists are few, this causes that the messenger is a science man. For example, Fermat's work over the theory of probabilities is known thanks to the correspondence that he had with Pascal, or the letters that sent Pascal to Carcavi in which he finds the length of arc of a cycloid, or the famous last theorem of Fermat that is known through a footnote in a manuscript \cite{history}.\\

It was the development of the press machine (mid XVth century) which guaranteed the diffusion of the knowledge, and at the same time gave an impulse to the intellectual production. For example, it was thanks to the press machine that in the first half of XVIIth century  Galileo Galilei could spread his manuscript  "Dialogue Concerning the Two Chief World Systems". It's interesting to remember that Galileo wrote his manuscripts in dialogue form, which allowed it to arrive at a not so specialized public. Soon in the second half of century XVII, Newton gave a strong impulse to modern science, at the same time the Royal Society created the present form to communicate and to validate the scientific knowledge: the journal, and during long time the journals have been the best form to communicate and to impel the scientific development.\\
|
Now, journals continue being means of diffusion and promotion of the scientific work with established rules; The published works must be original and no article can be published in two journals. In the same way the responsibility of an investigation is due to recognize by means of citations. In the publication process the articles must be analyzed and discussed by experts for their later publication, which allows to maintain a control on through the process. Nowadays we speak of 50,000 annual international publications in mathematics \cite{odlyzko3}.\\ 

Recently, in XXth century, with the invention of telecommunications and computers, online communications have appeared. In 1961 an experiment that involved the article dissemination based on preprints via IEGs (information Exchange Groups) was made. This service was supported by NIH (National Institutes of Health) in the USA, which counted with 2500 members and 80$\%$ of the memos were published in printed journals. The objective of this experiment was to provide immediate communication between investigators, but it came to an end in March of 1967 because IEG preprints is complete publications. This represented a danger for the existing journals since their use would be reduced. In addition IEG preprints violated the requirements of many journals since the submitted documents could not be placed elsewhere \cite{till}.\\

Later in August 1991 thanks to the efforts of Paul Ginsparg, a server for the centralized distribution of electronic preprints settled down, based on the National Laboratory at Los Alamos (LANL) \cite{lanl}, which counted with the support of APS (American Physics Society). The initial center was the Theoretical Physics of High Energies (Hep-th) and given its fast acceptance, it extended to many other areas of physics, the nonlinear mathematics, and computer sciences. Today's e-Print servers (initially called preprint server) are growing to new areas with more adepts and they are not being restricted, this is the case of arXiv which counts with 18 mirrors, 70$\%$ of the e-Prints there published are also in journals and 20$\%$ of them are conferences \cite{langer}.\\

\section {E-Print Servers}
Perhaps one of the most attractive factors of online publications (around 500 \cite{odlyzko3}) is that they allow a systematic consultation, then we can access in a selective form to the information using a search motor, which saves a lot of time when doing a consultation. Additionally the physical space in the libraries is reduced, of the same form we save hard copy printing and the time to obtain an article, in addition not always (unless we are suscribided to the specific journal) we can consult a foreign publications in printed. The slogan is: ``You just print what you use''. The costs of an electronic publication are low and its access is easy, for example a library (books, journals, etc.) in an institution can have a cost of 1/4 to 1/3 of the total of its actual expenses\cite{odlyzko1}. The e-Print servers are very used by the academic community by its low cost of maintenance and publication \cite{Ginsparg}, in addition the access to publications is free; arXiv is a good example of these. This is not something isolated within electronic media, e.g Linux, an operating system with free access that has more adepts, or the Napster phenomenon, which allowed free access to music; the impact of Google, given its efficiency in the search web and its little publicity, etc.\\

Our interests is in e-Print servers, which have generated a greater publicity, proliferation and intellectual interchange, maximizing the impact, speed and progress of scientific development, e.g. Maldacena's work [hep-th/9711200] based initially during 10 months in LANL, was one of the most cited in 1998 with 456 citations, or Polchinski's work  [hep-th/9611050] which has remained by more than 4 years as e-Print with 688 citations (see \cite{spires}). This shows the impact of works based on e-Print server.\\

When we want to establish a publication, one of the most severe problems is its prize, and considering the benefits before mentioned of electronic formatting, the publications have been forced to change or to maintain an online version to promote the intellectual production. This has entailed some journals to adapt quickly, among some: Physics Review Letters (PRL) \cite{prl}, The Journal of High Energy Physics (JHEP) \cite{jhep} (Pioneering in totally electronic journals of investigation in physics),  Mathematical Physics Electronic Journal (MPEJ) \cite{mpej}, Central PubMed \cite{pmed}, the last three of free access. On the other hand, costs are those which determine if a online publication is or not of free access, and since it's not possible to acces to all existing literature at nondeveloped  countries due to its costs, then the competitive investigation is restricted in them, where the theoretical work even can be competitive.\\

A determining factor for the promotion of the scientific work is perhaps the access to information, and the access to good quality information, which can be recognized by its innovation, rigor and its technological applications, or given the case by its pedagogical or epistemologic content. Access to information is easily obtained by electronic means, nowadays it's possible to maintain communication through email, e-Print server, web sites, chat or forums with experts, this without great difficulty. Nevertheless today electronic journals or e-Print servers have the same inherited primary characteristic of their precesores, they contain only text and images. Perhaps in a near future no only what we write will be fundamental, also will be important as it seams(remember the Galileo's work), the multimedia will be useful when the intention is to comunicate with a nonspecialized public\cite{Negroponte}.\\ 

On another side, when we review the statistics that correlate articles cited with their publication in journals free or not \cite{lawrence}, we find that the citation dynamics diminishes when the journals are not of free access. We present the annual citation dynamics of some of the most cited papers until year 2001 in High Energy Physics (HEP) over a database of $N_p=477359$ papers. These are: $\#$6 cited 2696 times, $\#$9 cited 2447 times, $\#$10 cited 2369 times, and $\#$11 cited 2309 times, according to SLAC Library (See Spires\cite{spires}).
\begin{figure}[ht]
\centering
\includegraphics[width=5.4cm, height=3.6cm]{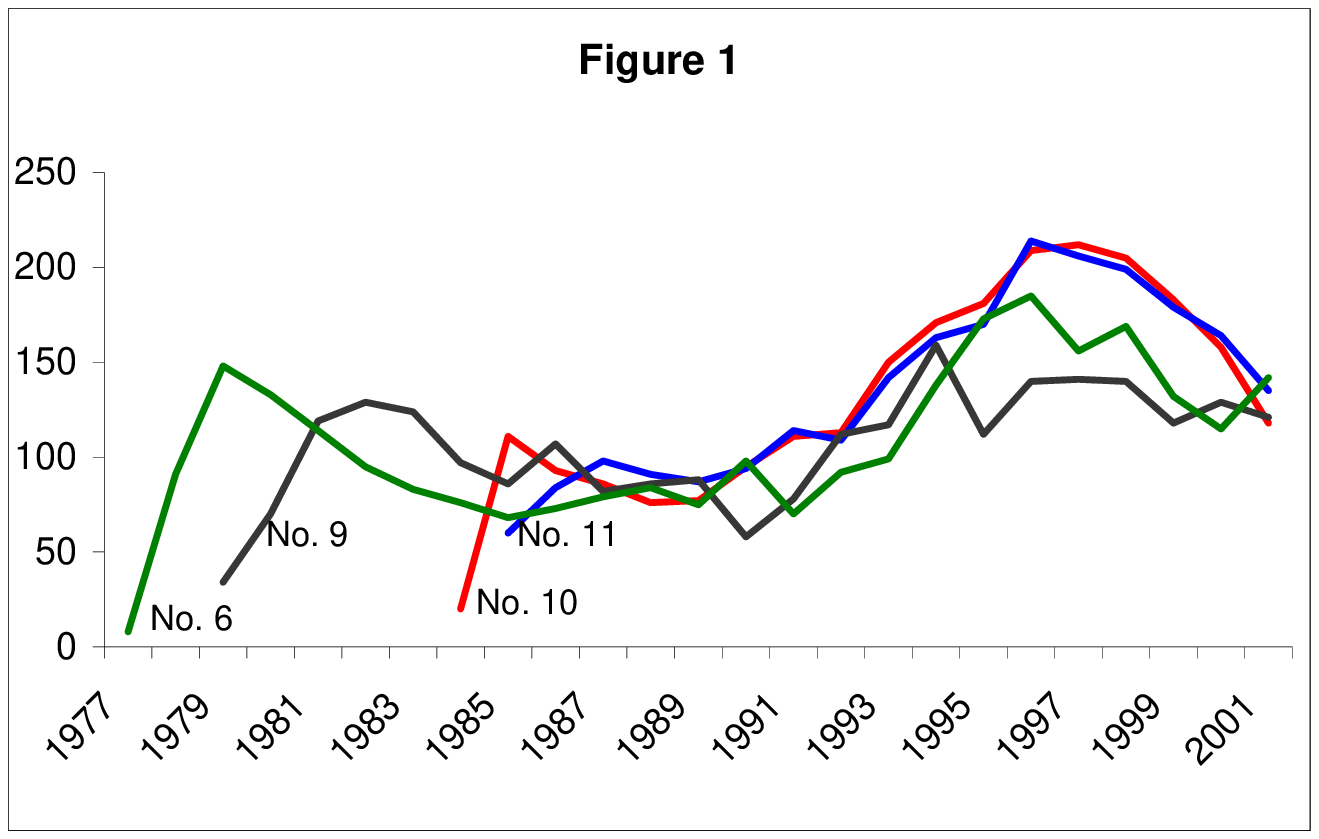}
\includegraphics[width=5.4cm, height=3.6cm]{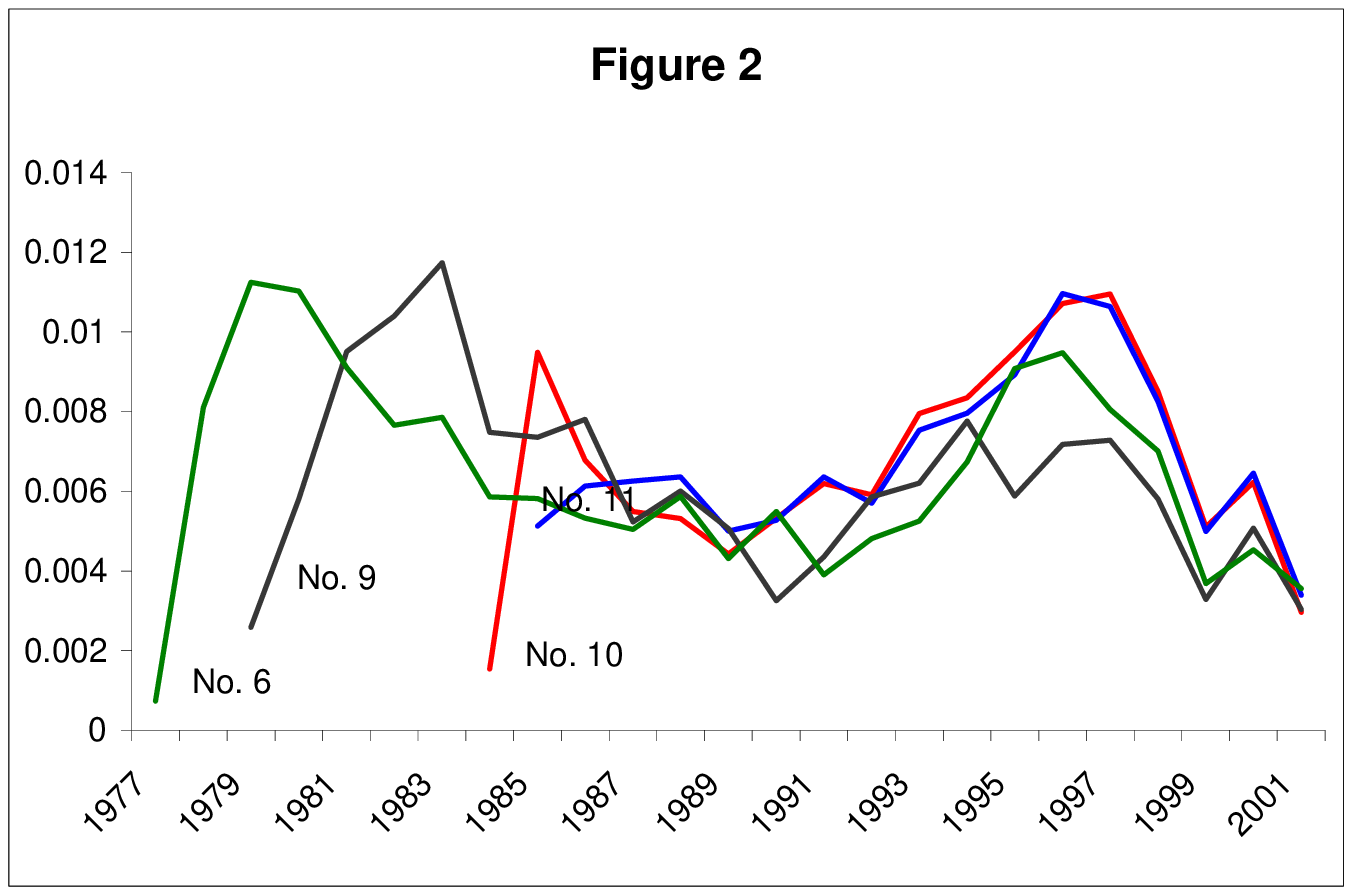}
\end{figure}
This database includes papers published in journals and e-Print servers and it can be observed that the citation dynamics for these articles changed after the year 1991, in which the Los Alamos preprint server was established. In figure 1, we show the impact measure of each paper along time ($I_{\# i}(y)$). In figure 2, we show $I_{\#i}(y)/N_p(y)$ (Here $I_{\# i}(y)$ represents the number of times the $\# i$ article has been cited in year $y$, and $N_p(y)$ represents the articles published in year $y$ and registered by the SLAC database). The dynamics shows that after some time it tends to zero in every case. Given the previous commentaries, the question of wether journals continue being a suitable format for the promotion and diffusion of the intellectual work or not. 

\section {A Possibility to Future Scientific Communications}
When a work is published by a journal, this means that an idea or an investigation has been registered, and that a group of experts endorses and supports the investigation. At the same time an investigation is being communicated to the scientific community (which implies to have access to this community), this is very well achieved by e-Print servers. On the other hand, if the citation dynamics is considered within the scientific community, it can be deduced that an article acquires relevance and status when its endorsed and cited in other works, and by other investigators; this is to say when it's useful to others. Therefore a work published as an e-Print acquires the same status of a publication done by a journal at the moment in which it's cited by someone different from the author(s), and the impact of e-Print would be moderated by it's number of citations and by those who cite the work (this last one is not novel and is something that always has been implicit within the social dynamics of the scientist community).\\ 

Although the process of selection in e-Print server is minimum, this system could work if a good culture of scientific communication is acquired \cite{Negroponte}, which would benefit to all. For example, one of the frequent vices of the scientific investigator is to publish by the mere fact of publishing (see also other commentaries in \cite{o'connell}). 
The same author must make his work valid and to select what is publishable and what is not. The investigator must remember that when he submits an article, his prestige is in game and this is decisive if he's interested in that his work is recognized, important when asking for support for investigation projects.\\

E-Print servers has been a system that recognizes the responsibility of a work, in which the very source of of documents (tex, txt, html, etc.) can be found. This fact (direct access to sources) is not matter of worry. The plagiarism of ideas or the presence of saboteurs (for example: to supplant a name, etc.) is something unseen and the scientists are communicating their ideas almost instantaneously, with the confidence of which its investigation is recognized (remember the works of Maldacena and Polchinski). Also, some recognized physicists who work in the area of High Energy Physics have begun to use just the e-Print archives and less the journals, today it's common to find citations to works that have been published only in an e-Print server.\\

We are giving a great deal of importance to citations, and it's possible to ask how to establish the status of an article according to its citations. The use of a directed graph can be a good method for the article classification (see \cite{ayala}), which is defined as follows: Given two articles represented by the nodes $A,B$, we say that $A$ is connected with $B$ ($A\to B$) if $A$ cites $B$. In addition we may consider the weight of $B$ as the number $i_{B}$ of connections to $B$, i.e. the number of times that $B$ is cited, in such a way that the citations made by the same group of authors or a sub-group of it are excluded. This graph can partially be ordered using its weights, since the smallest weight is zero and greater than some natural number $n\neq 0$. One says that two nodes $A,B$ are connected ($A\mapsto B$) if it's possible to find a way that it connects $A$ with $B$ (ej $A\to Q\to J\to M\to B$), and a node $C$ is essential (\^{C }) if it's a forced node of the $A\mapsto B$ way. A graph is connected if it does not have essential nodes, this is to say if the graph does not contain articles that connect isolated subjects.\\

Using the previous scheme of connected graphs or some other, we could measure the acceptance and the impact of e-Print, using the weight $i_x$ of each node, in such a way that e-Print acquires the status of a published article if it has a weight different from zero. On the other hand to take a statistic of the scientific production, is something that the electronic media can do without greater difficulty, this can be seen for example in Physics, Mathematics and Computer sciences the database of SPIRES-HEP \cite{spires} or Citeseer \cite{citeseer} respectively. 

At the moment, the journals will continue being a control mechanism and consultation and diffusion of works of good quality.

\section*{Acknowledgments}
The author was partially supported by the {\it Fundaci\'on Mazda para el Arte y la Ciencia}. He would like to express his gratitude to Alvaro O'byrne by its commentaries and suggestions.

\hrule width 14.cm
\vskip 2.mm
$\natural$: Published in http://ArXiv.org 

\end{document}